\documentclass[twocolumn,prl,showpacs]{revtex4-1}
\usepackage{graphicx}
\usepackage{amsmath}
\usepackage{amsfonts}
\usepackage{mathtools}
\usepackage{subfigure}

\begin{document}
\title{Universal minimal cost of coherent biochemical oscillations}

\author{Lukas Oberreiter$^1$, Udo Seifert$^1$, and Andre C. Barato$^2$}
\affiliation{$^1$ II. Institut f\"ur Theoretische Physik, Universit\"at Stuttgart, 70550 Stuttgart, Germany\\
$^2$ Department of Physics, University of Houston, Houston, Texas 77204, USA}

\parskip 1mm
\def\d{{\rm d}}
\def\Ps{{P_{\scriptscriptstyle \hspace{-0.3mm} s}}}
\def\MF{{\mbox{\tiny \rm \hspace{-0.3mm} MF}}}
\def\ts{\tau_{\textrm{sig}}}
\def\tos{\tau_{\textrm{osc}}}
\begin{abstract}
Biochemical clocks are essential for virtually all living systems. A biochemical clock that is isolated from an external periodic signal and subjected to fluctuations can oscillate 
coherently only for a finite number of oscillations. Furthermore, such an autonomous clock can oscillate only if it consumes free energy. What is the minimum amount of free energy
consumption required for a certain number of coherent oscillations? We conjecture a universal bound that answers this question. A system that oscillates coherently for $\mathcal{N}$ oscillations has 
a minimal free energy cost per oscillation of $4\pi^2\mathcal{N} k_B T$. Our bound is valid for general finite Markov processes, is conjectured based on extensive numerical evidence, 
is illustrated with numerical simulations of a known model for a biochemical oscillator, and applies to existing experimental data.       
\end{abstract}

\maketitle

Biological rhythms \cite{nova08,gold08,ferr11} are essential for the proper operation of living systems ranging from bacteria to 
mammals. The most prominent biological oscillations are circadian rhythms that take 24 hours, but the period
of biological rhythms has a wide range. One of the best understood circadian oscillators is KaiC, a protein 
that regulates the circadian clock of Cyanobacteria \cite{naka05,dong08,paji17}. In the presence of ATP and a few other 
molecules, KaiC molecules form a finite system of chemical reactions that produces circadian oscillations that can be observed in 
the laboratory. Another example of a finite system of chemical reactions that produces oscillations are  
synthetically engineered genetic circuits \cite{potv16}.              

In this context, consider an autonomous biochemical system composed of a finite number of molecules that oscillates. 
Fluctuations will eventually destroy the precision of the oscillations, i.e., if we consider different realizations 
of this biochemical system, they will dephase after some time. The quantity that characterizes the precision 
is the number of coherent oscillations $\mathcal{N}$ \cite{bark00,gasp02,gonz06,more07,risa07,eysm13}, which is the decay time divided by the period of 
oscillation, as observed in a time-correlation function. In other words, if we isolate a biochemical clock in a ``dark room'', i.e.,  without any external 
periodic stimuli, $\mathcal{N}$ roughly gives the number of oscillations for which the clock will still function properly. 

A biochemical clock consumes free energy, in the form of ATP hydrolysis, or some other chemical fuel. 
A fundamental question then is: What is the minimal cost to sustain a biochemical oscillator with 
a precision quantified by $\mathcal{N}$? This cost can be characterized by the entropy production of stochastic 
thermodynamics \cite{seif12}. 

Two studies quite related to this question are the following. First, in \cite{cao15}, the relation between 
the uncertainty in the period of oscillation, a quantity related to the number of coherent oscillations $\mathcal{N}$, 
and the entropy production was analyzed for several models. It was found that this uncertainty decreases linearly 
with the entropy production. Second,  in \cite{bara17a}, a thermodynamic bound on $\mathcal{N}$ has been conjectured. 
This bound is expressed in terms of the thermodynamic affinity that drives the system out of equilibrium. This thermodynamic 
affinity does not suffice to know the free energy the system dissipates, which is quantified by the entropy production that 
additionally depends on detailed kinetic parameters.

Several works have addressed the relation between biochemical oscillations and stochastic thermodynamics  
\cite{qian00,nguy18,fei18,wier18,mars19,zhan20,junc20a,junc20b,frit20}.
More generally, the relation between entropy production and precision of some kind has been a quite 
active area in the last 10 years in biophysics \cite{qian07,lan12,meht12,palo13,skog13,lang14,gove14a,bara14a,sart14,hart15,bo15,ito15,mcgr17,chiu19,rana20,sear21}.

In spite of all this research, the fundamental question raised above has remained unanswered. In this paper, we find 
that the number of coherent oscillations $\mathcal{N}$ has the  minimal cost 
\begin{equation}
\Delta S\ge 4\pi^2\mathcal{N},
\label{eqmainresult}  
\end{equation}
where $\Delta S$ is the average entropy production per oscillation. For instance, 
a biochemical oscillator with $10$ coherent oscillations must dissipate 
at least approximately $395 k_B T$ per oscillation, where $k_B$ is Boltzmann's constant and $T$ is the temperature.
This bound is independent of any details of the biochemical system. In fact, this bound can be violated  only for $\mathcal{N}< (2\pi)^{-1}$, which amounts 
to a vanishing, practically undetectable, number of coherent oscillations. Our main result in Eq. \eqref{eqmainresult} is a conjecture based on extensive numerical evidence.

   
A prominent relation in stochastic thermodynamics superficially related to our result is the thermodynamic uncertainty relation \cite{bara15a}. 
This relation establishes the minimal universal cost, as quantified by entropy production, related to the uncertainty of any thermodynamic current. 
Much work on the thermodynamic uncertainty relation has been done since its discovery \cite{ging16,piet16,nguy16,piet18,pole16,nyaw16,guio16,piet17b,horo17,pigo17,proe17,maes17,hyeo17,bisk17,bran18,nard18,chiu18,bara18c,dech18,caro19,liu19,guar19,koyu20,ito20,hase21}.
In particular, the relation between the uncertainty of a thermodynamic current and the number of coherent oscillations has been analyzed in \cite{nguy18}.
As shown in this reference the precision of biochemical oscillations is not well quantified by the uncertainty of a thermodynamic current.
Hence, the uncertainty of a thermodynamic current is a different mathematical object and it has a different physical interpretation as a quantifier of precision.
From the perspective of stochastic thermodynamics, the question we answer in this letter can be asked as follows. What is the 
minimal cost of the number of coherent oscillations as they are visible in time-correlation functions? Whereas the question answered by the thermodynamic uncertainty relation 
is:  What is the minimal cost of precision of a thermodynamic current? The bound conjectured in Eq. \eqref{eqmainresult} has the same degree of universality of 
the thermodynamic uncertainty relation first conjectured in \cite{bara15a}.

  
 We consider continuous-time Markov processes with a finite number of states $\Omega$. The transition rate from state $i$ to 
 state $j$ is denoted $k_{ij}$. The time evolution of the probability to be in state $i$ at time $t$, denoted $P_i(t)$, is determined 
 by the master equation 
\begin{equation}
\frac{d}{dt}P_i(t)=\sum_j\left[P_j(t)k_{ji}-P_i(t)k_{ij}\right].
\label{eqME}  
\end{equation}
In matrix form this equation can be written as $\frac{d}{dt}\mathbf{P}(t)=\mathbf{L}\mathbf{P}(t)$, where $\mathbf{P}(t)=\{P_1(t),P_2(t),\ldots,P_\Omega(t)\}^T$ and  
\begin{equation}
\mathbf{L}_{ij}\equiv k_{ji}-\delta_{i,j}\sum_l k_{il}
\label{eqL}  
\end{equation}
is the stochastic matrix. We assume that if $k_{ij}\neq 0$ then $k_{ji}\neq 0$. 
Such Markov processes are an appropriate mathematical framework to describe a system of chemical reactions.

The long-time stationary distribution is denoted by $P_i$. This distribution is the right eigenvector associated with the eigenvalue of $L$ with largest real part, which is $0$. 
The stationary rate of entropy production from stochastic thermodynamics is \cite{seif12} 
\begin{equation}
\sigma\equiv \sum_{ij}P_ik_{ij}\ln\frac{k_{ij}}{k_{ji}}\ge 0.
\label{eqsigma}  
\end{equation}
The rate of entropy production vanishes only if the steady state is an equilibrium one. An equilibrium steady state fulfills the detailed balance relation $P_ik_{ij}=P_jk_{ji}$
for all pairs $i,j$.

Consider a system of chemical reactions that generates oscillations and is well described as a Markov process. The concentration of an oscillating chemical species can generically be written 
as a time-correlation function, which is a linear combination of the time-dependent probabilities $P_i(t)$ that are obtained as the solution of Eq. 
\eqref{eqME}. Therefore, the period and decay time of the biochemical oscillations are quantified by the first non-trivial eigenvalue of $\mathbf{L}$, i.e., the non-zero eigenvalue with 
the smallest modulus of its real part \cite{bara17a}.

This eigenvalue is written as $\lambda= -\lambda_R\pm i\lambda_I$.  The decay time is $\lambda_R^{-1}$ and the period of oscillation is $2\pi\lambda_I^{-1}$. The
number of coherent oscillations is given by         
\begin{equation}
\mathcal{N}\equiv \frac{1}{2\pi} \frac{\lambda_I}{\lambda_R}.
\label{eqNdef}  
\end{equation}
This quantity $\mathcal{N}$ that quantifies the precision of the oscillations is dimensionless, while the entropy production rate $\sigma$ that quantifies the cost has dimension of $t^{-1}$. 
In order to make the cost dimensionless   we consider the entropy produced in one period of oscillation
\begin{equation}
\Delta S\equiv \sigma\times(2\pi\lambda_I^{-1}).
\label{eqdeltaS}  
\end{equation}


\begin{figure*}[t]
\subfigure[]{\includegraphics[width=72mm]{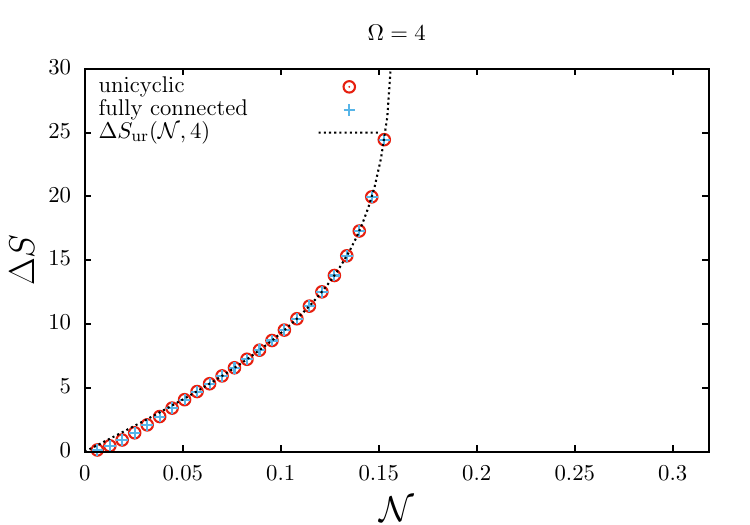}\label{fig1a}}\hspace{3mm}
\subfigure[]{\includegraphics[width=72mm]{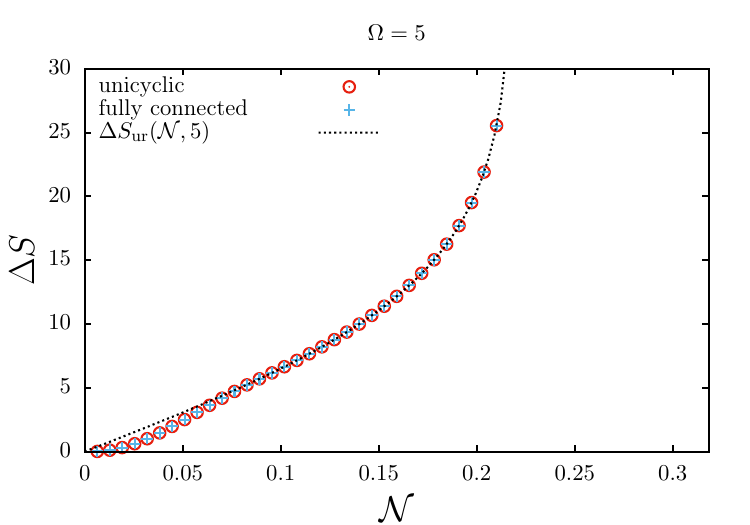}\label{fig1b}}\hspace{3mm}
\subfigure[]{\includegraphics[width=72mm]{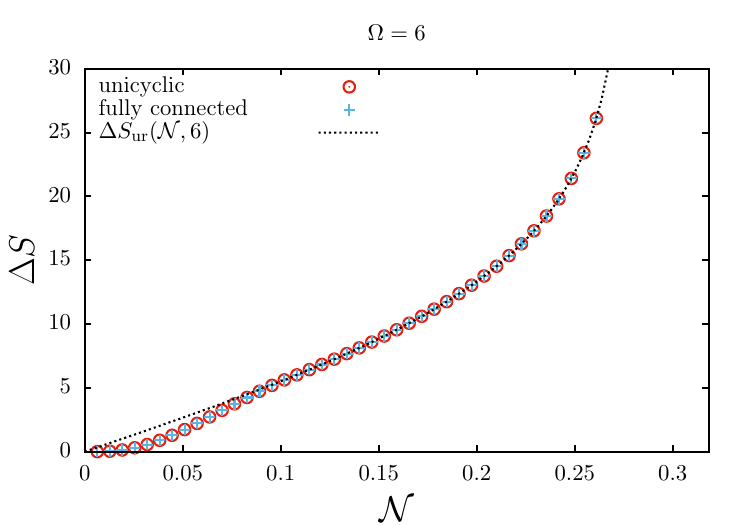}\label{fig1c}}\hspace{3mm}
\subfigure[]{\includegraphics[width=72mm]{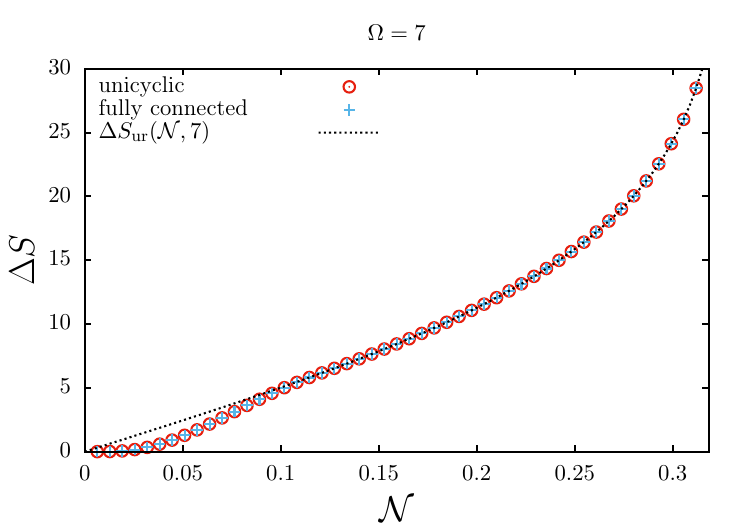}\label{fig1d}}
\vspace{-2mm}
\caption{(Color online) Minimal cost per period $\Delta S$ as a function of the number of coherent oscillations $\mathcal{N}$, obtained from numerical minimization for fully connected networks and for unicylic networks. Both are compared to the analytical form for uniform rates in Eq. \eqref{equr}.}   
\label{fig1} 
\end{figure*}

Unicyclic networks \cite{derr83} play a central role in our calculations. For these networks, transition rates from state $i$ are nonzero only to its two nearest neighbors $i+1$ and $i-1$. 
Unicyclic networks have periodic boundary conditions,  for $i=\Omega$ the right neighbor is $i+1=1$ and for $i=1$ the left neighbor is  $i-1=\Omega$. The transition rate from 
$i$ to $i+1$ is denoted by $k_{i,i+1}$. The affinity of a unicyclic network is defined as 
\begin{equation}
A\equiv \ln\left(\prod_{i=1}^\Omega\frac{k_{i,i+1}}{k_{i,i-1}}\right).
\label{eqAff}  
\end{equation}
The affinity $A$ is the thermodynamic force that drives the system out of equilibrium. The system reaches a stationary equilibrium state only if $A=0$.    

Consider a unicyclic network with affinity $A$ and number of states $\Omega$. Uniform rates are given by $k_{i,i+1}=k\textrm{e}^{A/\Omega}$ and 
$k_{i,i-1}=k$, for all $i$, where $k>0$ is a constant that sets the time scale of the rates. For such uniform rates, from the fact that 
the stationary probability of a random walk in a circle is uniform and from Eq. \eqref{eqsigma}, we obtain   
\begin{equation}
\Delta S_{\textrm{ur}}= \frac{2\pi A}{\Omega \sin(2\pi/\Omega)}.
\label{equni1}  
\end{equation}
From the calculation of the first non-trivial eigenvalue and Eq. \eqref{eqNdef}, we obtain 
\begin{equation}
\mathcal{N}= \frac{1}{2\pi} \cot\left(\frac{\pi}{\Omega}\right)\tanh\left(\frac{A}{2\Omega}\right) .
\label{equni2}  
\end{equation}
By eliminating the affinity $A$ we obtain, 
\begin{equation}
\Delta S_{\textrm{ur}}(\mathcal{N},\Omega)= \frac{4\pi}{\sin(2\pi/\Omega)}\textrm{arctanh}[2\pi\mathcal{N}\tan(\pi/\Omega)]\ge 4\pi^2\mathcal{N},
\label{equr}  
\end{equation}
where the inequality comes from the fact that $\Delta S_{\textrm{ur}}(\mathcal{N},\Omega)$ is 
a decreasing function of $\Omega$ and $\lim_{\Omega\to\infty}\Delta S_{\textrm{ur}}(\mathcal{N},\Omega)= 4\pi^2\mathcal{N}$. 
This analytical expression  will be important for the derivation of our main result.




Let us now consider the procedure of obtaining $\Delta S$ as a function of $\mathcal{N}$. Both quantities  depend on the transition rates. Numerically 
we can impose the constraint that $\mathcal{N}$ is fixed and minimize  $\Delta S$. Technically this minimization is done with the \texttt{Python} routine 
\texttt{scipy.optimize.minimize}. We have performed this minimization for $\Omega=4,5,6,7$, for both fully connected networks, 
which include all possible networks since some of the rates can be set to zero, and for the more restricted case of  unicyclic networks. As shown in Fig. \ref{fig1}, 
both procedures lead to the same minimum. For large enough $\mathcal{N}$ the minimum has the functional form that is obtained for the case of uniform rates in 
Eq. \eqref{equr}. For small $\mathcal{N}$ the numerical minimum goes below the expression in Eq. \eqref{equr}. The value of $\mathcal{N}$ for which the minimum deviates from the expression in 
Eq. \eqref{equr} gets larger for larger system size $\Omega$. Therefore, our numerical results show that the minimum of $\Delta S$ for fixed $\mathcal{N}$ 
is given by the numerical minimization of unicyclic networks.


\begin{figure}
\includegraphics[width=75mm]{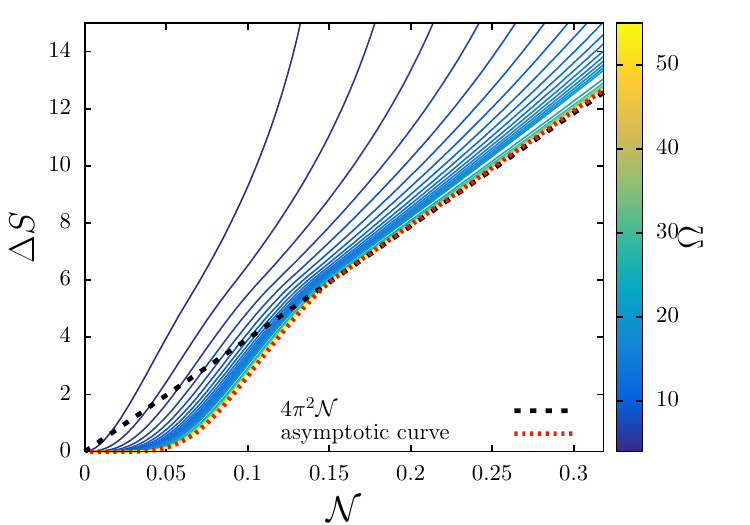}
\vspace{-2mm}
\caption{(Color online) Numerical minimization for unicycle networks from $\Omega=4$ up to $\Omega=55$. The red dashed line 
is the asymptotic curve $\Omega\to \infty$ obtained with numerical interpolation.}
\label{fig2} 
\end{figure}

Restricting to unicyclic networks allows us to calculate the numerical minimal $\Delta S$ for fixed $\mathcal{N}$ from $\Omega=4$ up to $\Omega=55$, as shown in Fig \ref{fig2}. From these 
curves we obtain the curve corresponding to the limit $\Omega\to \infty$. In particular, for fixed $\mathcal{N}$ we fit a function of the type
$\Delta S(\Omega)= a*\Omega^b + c$, where the parameter $c$ provides the limiting value of $\Delta S$. This limit curve gives our main result in Eq. \eqref{eqmainresult}. For $\mathcal{N}\ge (2\pi)^{-1}$,
the minimal cost $\Delta S$ for a certain number of coherent oscillations $\mathcal{N}$ is given by the $\Omega\to \infty$ function for uniform rates in Eq. \eqref{equr}. For $\mathcal{N}< (2\pi)^{-1}$,
the minimum $\Delta S$ goes below $4\pi^2 \mathcal{N}$ and its form is given by the numerically determined asymptotic curve in Fig. \ref{fig2}. Since, a system with $\mathcal{N}= (2\pi)^{-1}$
will not show any visible oscillation, for any biochemical oscillator with a reasonable number of coherent oscillations the bound in Eq. \eqref{equr} establishes the minimal universal cost of 
coherent oscillations. However, this small $\mathcal{N}$ region, for which uniform rates do not minimize $\Delta S$, is an important mathematical feature of the bound that should be taken 
into account in future analytical studies.


\begin{figure}
\includegraphics[width=75mm]{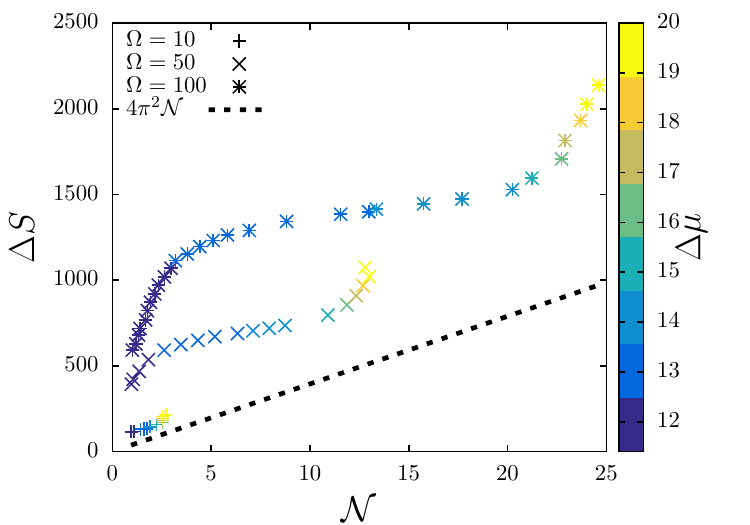}
\vspace{-2mm}
\caption{(Color online) Illustration of the bound with a model for a system of KaiC molecules. Data for $\mathcal{N}$ is taken from \cite{nguy18}. The parameter $\Omega$ 
represents the number of KaiC molecules and the parameter $\Delta \mu$ is the free energy of one ATP hydrolysis.}
\label{fig3} 
\end{figure}

In order to illustrate our bound we compare Eq. \eqref{eqmainresult} to an established  model for a system of KaiC molecules in Fig. \ref{fig3}. 
The data for $\mathcal{N}$ are taken from Fig. 11 in Ref. \cite{nguy18}. The definition of the model with the exact parameters 
are given in Sec. II C of this reference. The parameter $\Omega$, following the notation of this reference, is the number of KaiC molecules (and
not exactly the total number of states used in our calculations above). The parameter $\Delta \mu$ is the free energy of one ATP hydrolysis, i.e.,
the affinity that drives the system out of equilibrium. Each KaiC molecule has 6 phosphorylation sites and the quantity that oscillates is the phosphorylation 
level of the system of KaiC molecules. 

The cost $\Delta S$ for this more complex model stays above the lower bound conjectured here. For larger system sizes $\Omega$, $\Delta S$ is substantially above the bound.
Since both $\mathcal{N}$ and $\Delta S$ increase linearly with system size, it is not clear whether this observation is generic.    
Even for larger system sizes, the exact $\Delta S$ is not more than one order of magnitude above the bound. This particular example shows that a 
reasonably realistic model can operate close to the bound. 

Our bounds can be seen from a different perspective. Consider a biochemical oscillator for which we can measure a time-correlation function of 
an oscillating chemical concentration. From this correlation function, $\mathcal{N}$ can be evaluated, which together with Eq. \eqref{eqmainresult} 
implies a lower bound on the amount of free energy consumption per period. This lower bound is independent of any knowledge about the molecular details 
of the biochemical oscillator. In fact, our results can be applied to extant experimental data. As an example, we consider the repressilator from Ref. \cite{potv16}. 
We have estimated the number of coherent oscillations from the plots of the correlation function shown in  Fig. 1e and Fig. 2c in this reference, which correspond 
to two different situations. For the first case we estimate $\mathcal{N}\simeq 0.57$, which implies a minimum free energy cost per period of $22.5 k_B T$. For the second case we estimate
$\mathcal{N}\simeq 1.42$, which implies a minimum free energy cost per period of $56.0 k_B T$.  
       

The present results  are related to the bound conjectured in \cite{bara17a}. For simplicity, we will restrict the discussion to the case of unicyclic networks.  
In that reference it was shown  that for a network with affinity $A$ and number of states $\Omega$
\begin{equation}
\mathcal{N}\le \frac{1}{2\pi} \cot(\pi/\Omega)\tanh\left(\frac{A}{2\Omega}\right)\le \frac{1}{4\pi^2} A,
\label{eqoldbound}  
\end{equation}
where the second inequality corresponds to the limit $\Omega\to\infty$. Even though this bound and our result in Eq. \eqref{eqmainresult} look similar, they are mathematically and physically different. 
Mathematically, the bound in Eq. \eqref{eqoldbound} was obtained by maximizing $\mathcal{N}$ for fixed $A$ and $\Omega$ in \cite{bara17a}.  For this quantity, the maximum  is reached for uniform 
rates, independent of the value of $\mathcal{N}$. Our bound in Eq. \eqref{eqmainresult} can be obtained by maximizing the same function $\mathcal{N}$ but with the very different constraint that 
the entropy production per oscillation $\Delta S$ is fixed, which is analogous to minimize  $\Delta S$ with the constraint that $\mathcal{N}$ is fixed. Fixing $\Delta S$ is a much more 
complicated constraint due to its complex dependence on the transition rates. The present bound also displays the intricate nonlinear behavior for small $\mathcal{N}$,
in this regime $\mathcal{N}$ is not maximized for the case of uniform rates and Eq. \eqref{eqmainresult} is violated. The factor $4\pi^2$ shows up in both bounds since 
both are related to the limit $\Omega\to\infty$ for the case of uniform rates. 

The physical difference between both bounds comes from the physical difference between $A$ and $\Delta S$. The affinity $A$ is  independent of kinetic parameters and does not provide any 
quantitative information about free energy cost. The affinity $A$ depends on thermodynamic parameters such as ATP concentration and temperature. In contrast, entropy production per oscillation 
$\Delta S$ is a much more complex quantity. It depends on model specific kinetic parameters and quantifies the free energy cost per oscillation.


In conclusion, a biochemical oscillator can only show a certain number of coherent oscillations
if it dissipates a minimal amount of free energy, as dictated by Eq. \eqref{eqmainresult}. 
This bound constitutes a new universal law for biochemical oscillators. It is also an
inference tool: measurement of the decay time and period of oscillation of a 
biochemical clock leads to a lower bound on its free energy consumption.

From the perspective of stochastic thermodynamics, fluctuations cannot take any form but are bounded by
universal relations. Two of the most prominent such relation are the fluctuation theorem and the thermodynamic 
uncertainty relation. Our bound adds one more relation to the list, the number of coherent oscillations as visible in a time 
correlation function is bounded by the entropy production. 

The main result presented here is a conjecture based on numerical evidence, a full mathematical proof remains an open problem. 
From a broad perspective, it is interesting to ask whether the present bound and the thermodynamic uncertainty relation 
can be derived from a deeper master relation. Concerning applications, our universal bound could be used to, {\sl inter alia}, infer 
free energy dissipation and to optimize the number of coherent oscillations with a given free energy budget. Moreover, 
to investigate whether real biochemical clocks  operate close to the bound is an interesting question for future work.

\bibliographystyle{apsrev4-1}

\bibliography{/home/barato/work/papers/references/refs} 

\end{document}